\documentclass[showpacs,preprint,amsmath,amssymb]{revtex4}

\usepackage{graphicx}
\usepackage{dcolumn}
\usepackage{bm}

\begin{document}

%%%%%%%%%%%%%%%%%%%%%%%%%%%%%%%%%%%%%%%%%%%%%%%%%%%%%%%%%%%%%%%%%%%%%%%%%%%%%%%%%%%%%%%%%%%%
\title{Holonomy, Aharonov-Bohm effect and phonon scattering in superfluids}
%%%%%%%%%%%%%%%%%%%%%%%%%%%%%%%%%%%%%%%%%%%%%%%%%%%%%%%%%%%%%%%%%%%%%%%%%%%%%%%%%%%%%%%%%%%%
\author{Claudio Furtado and ~A. ~M. de ~M. Carvalho}

\affiliation{Departamento de F\'{\i}sica, CCEN,  Universidade Federal
da Para\'{\i}ba, Cidade Universit\'{a}ria, 58051-970 Jo\~ao Pessoa, PB, Brazil}

\author{~L.~C. Garcia de Andrade}

\affiliation{Departamento de F\'{\i}sica T\'eorica,
Instituto de F\'{\i}sica, Universidade Estadual do Rio de Janeiro.
Rua S\~ao Francisco Xavier 524, Maracan\~a, 20550, Rio de Janeiro, RJ, Brazil
}

\author{ Fernando Moraes}
\affiliation{Laborat\'orio de F\'{\i}sica Te\'orica e Computacional,
Departamento de F\'{\i}sica, Universidade Federal de Pernambuco,
50670-901 Recife, PE, Brazil}

\begin{abstract}
In this article we discuss the analogy between superfluids and a
spinning thick cosmic string. We use the geometrical approach to
obtain the geometrical phases for a phonon in the presence of a
vortex. We use loop variables for a geometric description of
Aharonov-Bohm effect in these systems. We use holonomy
transformations to characterize globally the ``space-time" of a
vortex and in this point of view we study the gravitational analog
of the Aharonov-Bohm effect in this system. We demonstrate that in the
general case the Aharonov-Bohm effect has a  contribution both from the
rotational and the  translational holonomy. We study also Berry´s
quantum phase for phonons in this systems.

\end{abstract}


\maketitle
%%%%%%%%%%%%%%%%%%%%%%%%%%%%%%%%%%%%%%%%%%%%%%%%%%%%%%%%%%%%%%%%%%%%%%%%%%%%%%%%%%%%%%%%%%%%%%%%%%%%%%%%%%%%%%%%%%%%%%%%%%%%%%%%%%%%%%%%%%%%%%%%%%%%%%%%%%%%%%%%%%%%%%%%%%%%%%%%%%%%%%%%
\section{Introduction}
%%%%%%%%%%%%%%%%%%%%%%%%%%%%%%%%%%%%%%%%%%%%%%%%%%%%%%%%%%%%%%%%%%%%%%%%%%%%%%%%%%%%%%%%%%%%%%%%%%%%%%%%%%%%%%%%%%%%%%%%%%%%%%%%%%%%%%%%%%%%%%%%%%%%%%%%%%%%%%%%%%%%%%%%%%%%%%%%%%%%%%%%%
The difficulty to test many of the cosmological models has been a
nuisance for physicists. In order to overcome this difficulty many
condensed matter physics systems have been extensively used as
laboratory for cosmological and gravitational
systems~\cite{bjp:moraes}. Such systems are known in the
literature as analogous systems. In recent years a  variety of
systems in condensed matter physics have been used as analog
models: Bose-Einstein condensates~\cite{prl:gar,pra:gar},
classical fluids
~\cite{prl:unruh,prd:unruh,prd:jac,prl:visser,cqg:visser} and
quantum fluids ~\cite{volo,boo}, moving dielectric media
~\cite{leon,brev}, non-linear electrodynamics, etc. The primordial
model was the sonic analogous one conceived by Unruh
~\cite{prd:unruh}. He starts from the  continuity and the Euler
equations for a classical fluid and got a geometric description
for the fluid equivalent to the  black hole solution. This system
has most of the known properties of a black hole, with the
advantage that its basic physics is completely known. The most
surprising result obtained by Unruh is that a non-relativistic
Newtonian fluid  propagating in a flat space plus time is governed
by the geometry of a Lorentzian $(3+1)$-dimensional curved space.
This formalism was obtained for quantum systems by
Volovik~\cite{volo}  to describe phonons in the presence of a
vortex. Volovik obtained a geometrical description for a series of
problems in superfluids and demonstrated that quantum fluids are
an excellent laboratory for some gravitational phenomena.

The superfluid Magnus force was defined by Hall and
Vinen~\cite{prs:hall} as a force between a vortex and a
superfluid. But, using a two-fluid hydrodynamics model  the Magnus
force  is not the only one acting on the  vortex transverse to its velocity.
There also exists another transverse force between the vortex and
quasiparticles  moving  with respect to the vortex.  For
phonons this force is called   Iordanskii force~\cite{jetp:ior}.  Sonin
\cite{jetp:sonin,prb:sonin}, Volovik ~\cite{volo}  and Stone
\cite{prb:stone}, on the other hand, have presented a detailed
review that shows that left/right asymmetry, in the scattering of
quasiparticles by the vortex line, arises from the fluid analogue
of the Aharonov-Bohm effect ~\cite{ejp:berry},  this effect gives
origin to  the Iordanskii force. An exact  expression  for
the transverse  force acting on a quantized vortex moving in a
neutral superfluid has been found recently\cite{prl:ao,prl:ao1,prl:wex,prb:wex}. In reference~\cite{prl:ao1}, Thouless, Ao and Niu  found  no Iordanskii
force linear in normal fluid density, in the fluid circulation,
and in the vortex velocity relative to the normal fluid component, in contrast  with the
previous results that got Iordanskii~\cite{jetp:sonin,prb:sonin} force. But this discrepancy was
solved in the paper by Thouless, Vinen, Geller, Fortin and Rhee
~\cite{prb:rhe} and Sonin~\cite{condm:sonin}.

 In this work we have used the analogue model of the
superfluid condensate constructed by Volovik to study from the geometrical point of view the
analogue of gravitational Aharonov-Bohm effect in these systems.
Volovik used the Landau theory of superfluids to show
that the energy of a quasiparticle moving in the
superfluid velocity field $\mathbf {v}_{s}(\mathbf{r})$,
which in the case of phonons has the spectrum
given by: $\epsilon({\mathbf p})=cp$,
is related to the momentum by the expression
\begin{equation}
\left(E-\mathbf{p}\cdot \mathbf{v}_{s}\right)^2=c^2p^2.
\end{equation}
The above equation can be written in a Lorentzian form with
$p_{\mu}=(E,\mathbf{p})$, thereby he wrote a metric tensor for
this system. We also note that the dynamics of phonons in the
presence of the velocity field is the same as the dynamics of
photons in the gravity field. In the work of Stone, he
investigated the scattering of phonons by a vortex moving with
respect to a superfluid condensate. They also study the analogy
between the Iordanskii force and the Aharonov-Bohm
effect~\cite{pr:aha}.

In a metric theory of gravitation, a gravitational field is
frequently related to a nonvanishing Riemann curvature tensor.
However, the presence of localized curvature can have effects on
geodesic motion and parallel transport in regions where the
curvature vanishes. The best known example of this nonlocal effect
is provided when a particle is transported around an idealized
cosmic string along a closed curve.The presence of the  string is
noticed by the particle even though there is no curvature along the trajectory. This situation corresponds to the gravitational
analogue~\cite {jphys:vile,prd:val,an:val} of the electromagnetic
Aharonov-Bohm effect~\cite{pr:aha}. These effects are of
topological origin rather than local. The electromagnetic
Aharonov-Bohm effect represents a global anholonomy associated
with the electromagnetic gauge potentials. Its gravitational
counterpart may be viewed as a manifestation of nontrivial
topology of space time. It is worth to call attention to the fact
that differently from the electromagnetic Aharonov-Bohm effect
which is essentially a quantum effect, the gravitational analogue
appears also at a  purely classical context. Thus, in summary, the
gravitational analogue of the electromagnetic Aharonov-Bohm effect
is the following: particles constrained to move in a region where
the Riemann curvature tensor vanishes may exhibit a gravitational
effect arising from a region of nonzero curvature from which they
are excluded. This effect may be viewed as a manifestation of the
nontrivial topology of space-time. In a more general sense,
particles constrained to move in a region where the Riemann
curvature is nonzero, but does not depend on certain parameters
such as velocity,  like in the case of moving mass
currents~\cite{nuo:law},  or the angular momentum of a rotating
body~\cite{nuo:fro}, in both examples in the weak field
approximation, may exhibit gravitational effects associated with
each one of these parameters in the respective cases. This kind of
gravitational effect we are calling generalized gravitational
Aharonov-Bohm effect.

 The analogous  of
gravitational Aharonov-Bohm effect recently was studied for
phonons in superfluids in the presence of a vortex. These  studies
consider the limit of distance far from  the vortex. In this
limit the space-time that describes the vortex is locally flat,
and guaranteeing therefore the analogy of the dynamics of
phonons in superfluids  and the quantum dynamics of particles
without mass in the presence of a cosmic string, where it appears as 
manifestation of the  gravitational Aharonov-Bohm effect. In the point
of view of an analogous model a vortex in a superfluid is
described for a curved space-time endowed with a metric of the
Painlev\'e-Gullstrand  type. In the limit far from the vortex, this
metric  has as limit the metric  that describes a spinning cosmic
string. Recently Fischer and Visser~\cite{prl:visfis,an:visfis}
 considered the acoustic propagation in the  presence of a
vortex and  studied the properties of  the sound waves in
this acoustic geometry. Then showed that the  metric of the vortex
differs strongly from the  metric of a spinning cosmic string for
 near  and  intermediate distances from the vortex  core. As is well
known, the metric  that describes the cosmic string, without
internal structure, is flat at any distance from the defect. In
contrast, the metric of the  vortex, for intermediate distance from it,
is not flat. If we consider the exact metric  that  describes the
vortex we need to use tools that consider the curved nature of the
space-time that describes the vortex. We will use the  calculation of
holonomy to study influences of the curvature in the dynamics of
phonons in the  vortex background, in this way,  all  effects of
space-time generated by the vortex, not only in the limit of large
distance from the vortex core, are considered.

 In this work we show that far from the vortex,
the approximation used in superfluids by Volovik leads
to a particular case of our metric which is obtained
by a local coordinate transformation similar to
a Lorentz rotation. We use the geometrical formalism
to study the sprouting of Berry phases related to this problem.
We also calculate the holonomy associated with this problem.
%%%%%%%%%%%%%%%%%%%%%%%%%%%%%%%%%%%%%%%%%%%%%%%%%%%%%%%%%%%%%%%%%%%%%%%%%%%%%%%%%%%%%%%%%%%%%%%%%%%%%%%%%%%%%%%%%%%%%%%%%%%%%%%%%%%%%%%%%%%%%%%%%%%%%%%%%%%%%%%%%%%%%%%%%%%%%%%%%%%%%%%%%
\section{Acoustical Line Element}
%%%%%%%%%%%%%%%%%%%%%%%%%%%%%%%%%%%%%%%%%%%%%%%%%%%%%%%%%%%%%%%%%%%%%%%%%%%%%%%%%%%%%%%%%%%%%%%%%%%%%%%%%%%%%%%%%%%%%%%%%%%%%%%%%%%%%%%%%%%%%%%%%%%%%%%%%%%%%%%%%%%%%%%%%%%%%%%%%%%%%%%%%

In this section we describe geometrically phonons propagating in
the presence of a vortex superfluid. We adopt the geometric
formulation for this problem given by Volovik. The dynamics of
phonons propagating in the velocity field of the quantized vortex
in the Bose superfluid $^{4}He$ is determined by the line
element~\cite{volo}
\begin{equation}
\label{phonon}
ds^2=\left(1-\frac{v^{2}_{s}}{c^2}\right)
\left(dt+\frac{N k d\phi}{2\pi (c^{2}-v^{2}_{s})}\right)^{2}
-\frac{dr^2}{c^2}-\frac{r^2}{c^2}d\phi^{2}-\frac{dz^2}{c^2},
\end{equation}
where $\vec{v}_{s}=N k \hat{\phi}/2\pi r$ is the velocity field
around the quantized vortices, $k$ is the quantum of circulation
and at last, $N$ is the circulation quantum number. Volovik used
the approximation that the metric ~(\ref{phonon}) is far from the
vortex, i.e. $v_{s}^{2}/c^{2}\ll 1$, to obtain the cosmic spinning
metric. We show that this assumption is not necessary and that a
simple global coordinate transformation is sufficient. We know
that for long distances of the vortex the metric  of
Painlev\'e-Gullstrand  has as  limit the metric  of a spinning
cosmic string. We use a coordinate transformation in the metric of
Painlev\'e-Gullstrand for transforming into  the space-time that has
the same form of a thick cosmic string and that we can better
compare our results with of the spinning cosmic string. Applying
the local coordinate transformation given by
\begin{subequations}
\label{global}
\begin{eqnarray}
d\phi'&=&d\phi \mbox{,}\\
z'&=&\frac{z}{c} \mbox{,}\\
r'&=&\frac{r}{c}  \mbox{,} \\
dt'&=&\sqrt{1-\frac{v^{2}_{s}}{c^{2}}}dt \mbox{,}
\end{eqnarray}
\end{subequations}
we obtain the metric analogous to the thick spinning cosmic
string~\cite{pla:anandan,cqg:solen} which is given by
\begin{equation}
\label{spinning}
ds^2=\left[dt'+\beta(r')d\phi'\right]^2-dr'^2-\alpha(r')^2r'^2d\phi'^2-dz'^2,
\end{equation}
where the functions $\alpha(r')$ and $ \beta(r')$,
are given respectively by
\begin{subequations}\label{trans1}
\label{functions}
\begin{eqnarray}
\alpha(r')&=&\sqrt{1-\frac{v^{2}_{s}}{c^{2}}}  \mbox{,}\\
\beta(r')&=&\frac{Nk}{c 2\pi}\left( 1-v^{2}_{s}{c^{2}}\right)^{-1/2}  \mbox{.}
\end{eqnarray}
\end{subequations}
The transformations ~(\ref{global}) are similar to the Lorentz
transformations of Special Relativity. Making an analogy with the
gravitational case, we note that the  function $\beta$ is
associated with the term of angular momentum by $\beta=-4\pi G J$.
We conclude that Volovik's acoustical metric has its form  similar
to the thick spinning cosmic string metric. In this way, this
metric includes the core structure of the vortex.

Note that when we are very far from the vortex the coordinate $r$ goes to infinity and we obtain
\begin{equation}
ds^{2}=(dt + \beta d\phi)^{2} - \alpha^{2}dr'^{2} - r'^{2}d\phi'^{2} - dz^{2},
\end{equation}
where now $\alpha$ and $\beta$ are constants and we obtain the
usual conical defect of the spinning cosmic string. This torsion
string is similar to the one obtained by Volovik  ~\cite{boo}
where analogy with point like torsion defects of Einstein-Cartan
Gravity ~\cite{cqg:garcia} play the role of Abrikosov vortices in
superconductors.
%%%%%%%%%%%%%%%%%%%%%%%%%%%%%%%%%%%%%%%%%%%%%%%%%%%%%%%%%%%%%%%%%%%%%%%%%%%%%%%%%%%%%%%%%%%%%%%%%%%%%%%%%%%%%%%%%%%%%%%%%%%%%%%%%%%%%%%%%%%%%%%%%%%%%%%%%%%%%%%%%%%%%%%%%%%%%%%%%%%%%%%%
\section{Holonomy in the Acoustical Metric}
%%%%%%%%%%%%%%%%%%%%%%%%%%%%%%%%%%%%%%%%%%%%%%%%%%%%%%%%%%%%%%%%%%%%%%%%%%%%%%%%%%%%%%%%%%%%%%%%%%%%%%%%%%%%%%%%%%%%%%%%%%%%%%%%%%%%%%%%%%%%%%%%%%%%%%%%%%%%%%%%%%%%%%%%%%%%%%%%%%%%%%%%
In this section we determine the holonomy
associated with the parallel transport of vectors
along closed curves around the vortices.
The holonomy can be used as a global classification of
differents kinds of spaces.
Hence it is very important for a topological description of these spaces.
Holonomy is employed in several areas of physics. Mathematically speaking, holonomy
are matrices that represent the parallel
transport of vectors, spinors, tensors, etc. This matrix provides information
on the curvature and topology of a given manifold.  The holonomy matrix can
be written as
\begin{equation}
\label{phase}
U_{AB}(C)={\cal P}\left(
-\int_{A}^{B} \Gamma_{\mu}(x(\lambda))\frac{dx^{\mu}}{d\lambda}d\lambda
\right),
\end{equation}
where $\Gamma_{\mu}$ is the tetradic connection and $A$ and $B$ are the
initial and final points of the path. Then, associated with every path
$C$ from a point $A$ to a point $B$,
we have a loop variable $ U_{AB}$ given by ~(\ref{phase}) which, by
construction, is a function of the path $C$ as a geometrical object.
It exists two types of holonomy: rotational or linear  holonomy that describes the
rotation of the  orthonormal frame parallel-propagated around a closed contour. That is
trivial for a family of metrics of three parameters  and of this form it does not
make  distinction between then and Minkowski's metric. This type of
problem is decided with  translational or affine holonomy that, in general,
distinguishes  those metrics. In general relativity the term affine
is usually followed by the appearance of torsion. Here we will use a description
for the  metric of the vortex in a theory of Einstein-Cartan.
Initially we will calculate rotational holonomy  and afterwards we will
calculate  translational holonomy.
The holonomy yields the topological information about the scattering
of phonons by the vortex. Transformations of holonomy make possible the study of the regions next,  intermediate and distant of the vortex. We can use the information given by the holonomy
calculus necessary to investigate the Aharonov-Bohm effect in
 phonon scattering. Notice, in the present case, the denomination Aharonov-Bohm effect  applies strictly only when we consider the limit $r>>1$.
For  other limits the space-time is curved and in  this form the analogy is not valid, therefore the geodesic movement of the  quasiparticle is affected by the local curvature of the space-time. We will use this nomenclature here having in mind this comment.
We can describe the metric (\ref{spinning}) in terms of
Cartan 1-forms basis as
\begin{subequations}
\label{basis}
\begin{eqnarray}
e^0&=& dt  + \beta  \mbox{,}          \\
e^1&=& \cos \phi dr-\alpha r \sin \phi d\phi     \mbox{,} \\
e^2&=& \sin \phi dr+\alpha r \cos \phi d\phi     \mbox{,} \\
e^3&=&dz \mbox{,}
\end{eqnarray}
\end{subequations}
where we have taken $\beta=0$. Initially we calculate the rotational holonomy,
in this case the
torsion or rotation does not
contribute to the holonomy. The connection forms
can be obtained
from the first Cartan structure equation,
which is given by
\begin{equation}
de^{a}+\omega^{a}_{b}\wedge e^{b}=0
\end{equation}
The connections $\omega^{a}_{b}$ are related to the spin
connection $\Gamma_{b}$ through the expression $\Gamma_{b}=
\omega^{a}_{b}dx^{b}$. Using the basis ~(\ref{basis}), the unique
non-null spin connection is the matrix
\begin{eqnarray}
\Gamma_{\phi}=\left(
\begin{array}{cccc}
1 &  0 & 0 & 0 \\
0 &  0 & (1-\alpha-\frac{1}{\alpha}\frac{v_{s}^{2}}{c^{2}}) & 0  \\
0 &  -(1-\alpha-\frac{1}{\alpha}\frac{v_{s}^{2}}{c^{2}}) &  0 & 0 \\
0 & 0 & 0 & 1
\end{array}\right) .
\end{eqnarray}
The holonomy associated with the parallel transport of vectors
around a closed curve  $\gamma$ is given by
\begin{equation}
\label{integral} U(\gamma)={\cal P}\exp
\left(-\oint_{\gamma}\Gamma_{\mu}dx^{\mu}\right).
\end{equation}
The curve $\gamma$ is taken as a circle or radius $r$, centered in
the origin. Thus, the integral can be written as
\begin{equation}
U(\gamma)=\exp\left[ -2\pi i(1-\alpha-\frac{1}{\alpha}\frac{v_{s}^{2}}{c^{2}})J_{12}\right].
\end{equation}
Or, in a matricial form, as
\begin{eqnarray}\label{AB}
U(\gamma)=\left(
\begin{array}{cccc}
1 &  0 & 0 & 0 \\
0 & \cos[2\pi(1-\alpha\frac{1}{\alpha}-\frac{v_{s}^{2}}{c^{2}})] &  \sin[2\pi(1-\alpha-\frac{1}{\alpha}\frac{v_{s}^{2}}{c^{2}})]& 0  \\
0 &  -\sin[2\pi(1-\alpha-\frac{1}{\alpha}\frac{v_{s}^{2}}{c^{2}})] &  \cos[2\pi(1-\alpha\frac{1}{\alpha}-\frac{v_{s}^{2}}{c^{2}})] & 0 \\
0 & 0 & 0 & 1
\end{array}\right) .
\end{eqnarray}
The matrix $J_{12}$ is the rotation generator around
the $z$-axis and is  the generator of the Lorentz  group.
This result means that when a vector is transported along
a region of non-null curvature the final vector is
changed. This is because of the existence of curvature
in the center of the vortex. This effect is similar to
the Aharonov-Bohm effect. Note that (\ref{AB}) gives a similar result for the
 holonomy transformation in the background of the thick cosmic string \cite{prd:val}. The crucial difference occurs in the fact that the space-time exterior to the cosmic string has Riemann tensor null contrary to the vortex case that has curvature in the exterior region. Far
from the vortex, where $\frac{v_{s}^{2}}{c^{2}}$ is small and can be neglected and  the holonomy
matrix is trivial,  we obtain no Aharonov-Bohm effects of rotational holonomy. In this way we need
to calculate the translational holonomy that gives non-trivial contributions  in the space-time that
contains  torsion or rotation\cite{petit,cqg:tod}.

To include the effects due to torsion,
we need to set $\beta$ non-null, rewrite the 1-form basis
with this term and then evaluate the holonomy. But
when we do that no effect appears due the presence of torsion.
This occurs because the holonomy that we are using
is not the appropriate one. We should use  a more
general holonomy known as translational holonomy \cite{petit,cqg:tod}.
The usual holonomy belongs to the Lorentz group
while the translational one to the Poincar\'e group.
Therefore we are going to rewrite
the metric (\ref{spinning}) as a Minkowski
metric through the following change of variables
\begin{subequations}
\label{trans1}
\begin{eqnarray}
t&=& T-\beta \phi          \mbox{,} \\
\phi&=& \theta / \alpha   \mbox{,} \\
z&=& Z                    \mbox{,} \\
r&=& R                \mbox{.}
\end{eqnarray}
\end{subequations}
The set of transformations (\ref{trans1}) could be written as a
product of homogeneous matrices\cite{an:val}. This way we need a
representation of the  holonomy group that has information of the
group of Lorentz and the group of the translations in
(3+1)-dimensions. Observing the  change of variables~(\ref{trans1}) we
put  this in  the   form of a homogeneous matrix
multiplication by the following procedure: let $M^B_A$ be a
five-dimensional matrix, with $A$ and $B$ running from 0 to 4. We
take $M^\mu_\nu$ equal to the rotation matrix given by
eq(\ref{AB}), $M^{3}_{4}= 2\pi \beta$. Thereby the points
$(t',\vec{x}')$ and $(t, \vec{x})$ are connected by
\begin{eqnarray}
\left(
\begin{array}{c}
t'\\
x'\\
y'\\
z'\\
1\\
\end{array}\right)
=\exp{(2\pi i \beta)T_{0}}\exp{(-i 2\pi(1-\alpha-\frac{1}{\alpha}\frac{v_{s}^{2}}{c^{2}}) J_{12})}\left(
\begin{array}{c}
t\\
x\\
y\\
z\\
1\\
\end{array}\right) .
\end{eqnarray}
were $T_{0}$ is given by
\begin{eqnarray}
\label{holonomiat}
T_{0} =\left(
\begin{array}{ccccc}
0 &  0 & 0 & 0 & 0 \\
0 &  0 & 0 & 0 & 0 \\
0 &  0&  0 & 0 & 0\\
0 & 0 & 0 & 0 & 1\\
0 & 0 & 0 & 0 & 0\\
\end{array}\right) .
\end{eqnarray}

The translational holonomy matrix is given by
\begin{eqnarray}
\label{holonomia}
U(\gamma)=\left(
\begin{array}{ccccc}
1 &  0 & 0 & 0 & 0 \\
0 & \cos[2\pi(1-\alpha-\frac{1}{\alpha}\frac{v_{s}^{2}}{c^{2}})] &  \sin[2\pi(1-\alpha-\frac{1}{\alpha}\frac{v_{s}^{2}}{c^{2}})]& 0 & 0 \\
0 &  -\sin[2\pi(1-\alpha-\frac{1}{\alpha}\frac{v_{s}^{2}}{c^{2}})] &  \cos[2\pi(1-\alpha-\frac{1}{\alpha}\frac{v_{s}^{2}}{c^{2}})] & 0 & 0\\
0 & 0 & 0 & 1 & 2\pi \beta\\
0 & 0 & 0 & 0 & 1\\
\end{array}\right) .
\end{eqnarray}
This matrix describes the behavior of a vector
 when it is parallel-transported around a
topological defect. In this analogous model the phonons see the
vortex as a non-Euclidean metric. In this case, the holonomy
provides the wave function when it is parallel transported around
the vortex. In this way the phase acquired by the wave function of
the phonons is given by the holonomy in the vortex background.
This is a manifestation of the Aharonov-Bohm effect for phonon
dynamics  in the presence of a superfluid  vortex.  In the limit
were  $ \frac{v_{s}^{2}}{c^{2}}$ is small, the expression
(\ref{holonomia}) is similar to the matrix holonomy for a thick
massless cosmic string were $\alpha=1$~\cite{prd:bur}.  All
contributions to the Aharonov-Bohm effect, in this limit, are
given by the translational holonomy, that has the following form
\begin{eqnarray}
\label{tholonomia}
U(\gamma)=\left(
\begin{array}{ccccc}
1 &  0 & 0 & 0 & 0 \\
0 & 1&  0& 0 & 0 \\
0 &  0 & 1 & 0 & 0\\
0 & 0 & 0 & 1 & 2\pi \beta\\
0 & 0 & 0 & 0 & 1\\
\end{array}\right) ,
\end{eqnarray}
where in this limit $\beta=\frac{Nk}{2\pi c}$. In this way the phase acquired
in the parallel transport far from the vortex  is given by
\begin{eqnarray}
  \label{trans}
  U(c)=\exp\{i \frac{Nk}{c }T_{0}\}.
\end{eqnarray}
Note that the holonomy transformation  in the vortex background  in the Volovik
analogue model is similar to the transformation of holonomy in the thick cosmic
string space-time and, in the limit $\frac{v_{s}^{2}}{c^{2}}<< 1$
has the same form of the holonomy transformation  for a massless thick cosmic string.
%%%%%%%%%%%%%%%%%%%%%%%%%%%%%%%%%%%%%%%%%%%%%%%%%%%%%%%%%%%%%%%%%%%%%%%%%%%%%%%%%%%%%%%%%%%%%%%%%%%%%%%%%%%%%%%%%%%%%%%%%%%%%%%%%%%%%%%%%%%%%%%%%%%%%%%%%%%%%%%%%%%%%%%%%%%%%%%%%%%%%%%%
\section{Geometric Properties of the Vortex Background}
%%%%%%%%%%%%%%%%%%%%%%%%%%%%%%%%%%%%%%%%%%%%%%%%%%%%%%%%%%%%%%%%%%%%%%%%%%%%%%%%%%%%%%%%%%%%%%%%%%%%%%%%%%%%%%%%%%%%%%%%%%%%%%%%%%%%%%%%%%%%%%%%%%%%%%%%%%%%%%%%%%%%%%%%%%%%%%%%%%%%%%%%
In this section we analyze some geometric quantities associated
with the vortex background. We can evaluate
the quantum torsion flux in the superfluid as an holonomy integral,
we consider a loop of constant  radius $r$, centered
in the origin
\begin{equation}
\label{torsion-flux}
\int_{\Sigma}Q^0=\oint e^0_{\gamma} =\oint \beta d\phi '=4G\oint Jd\phi ' =
\frac{8\pi}{\sqrt{1-v^2/c^2}}Nk,
\end{equation}
where $\Sigma$ denotes the surface where the flux is evaluated.
The time translation must correspond to torsion\cite{cqg:tod,pla:anandan,pla:kohler} being
nonzero inside the vortex string.
In the case of the vortex in $^4He$ superfluid
$k=\pi h/m_{4}$, where $m_{4}$ is the mass of
$^4He$ atom. So, the above expression can be
rewritten as
\begin{equation}
\int_{\Sigma}Q^0=\frac{N h}{2 m_4 c^2 \sqrt{1-v^2/c^2}}.
\end{equation}
In the thin
spinning cosmic string approximation one
obtains for the torsion quantized flux
\begin{equation}
\label{torsion-flux1}
\int_{\Sigma} Q^{0}= \frac{Nh}{2m_{4} c^{2}}
\end{equation}
The results given by equation~(\ref{torsion-flux1})
is analogous to the  the Bohr-Sommerfeld quantization integral. Thus the torsion
flux is naturally quantized. Notice that in the far from the vortex limit this is independent of the radius.

Now we consider the curvature two-form for the vortex background. Here,
also,  we consider a loop of constant  radius $r$, centered in the
origin. Making an analogy with the Anandan ~\cite{pla:anandan}
work we can write a Gauss-Bonnet theorem for the curvature flux of
the spinning cosmic string as
\begin{equation}
\oint_{\Sigma} R_{r\phi 2}^{2} dr \wedge d\phi = 2\pi(1-\alpha-\frac{1}{\alpha}\frac{v_{s}^{2}}{c^{2}}).
\end{equation}
and in the thin
spinning cosmic string approximation one
obtains
\begin{equation}
\int_{\Sigma} R_{r \phi 2}^{1} dr\wedge d\phi= 2\pi c^{2}
\end{equation}
for the curvature flux which is proportional to the sound speed $c$,  which we have considered
constant. This space-time can be compared  with the  generalized cosmic string studied by Vickers\cite{cqg:vic}.
%%%%%%%%%%%%%%%%%%%%%%%%%%%%%%%%%%%%%%%%%%%%%%%%%%%%%%%%%%%%%%%%%%%%%%%%%%%%%%%%%%%%%%%%%%%%%%%%%%%%%%%%%%%%%%%%%%%%%%%%%%%%%%%%%%%%%%%%%%%%%%%%%%%%%%%%%%%%%%%%%%%%%%%%%%%%%%%%%%%%%%%%
\section{Berry's Quantum Phase in Phonons in Superfluid}
%%%%%%%%%%%%%%%%%%%%%%%%%%%%%%%%%%%%%%%%%%%%%%%%%%%%%%%%%%%%%%%%%%%%%%%%%%%%%%%%%%%%%%%%%%%%%%%%%%%%%%%%%%%%%%%%%%%%%%%%%%%%%%%%%%%%%%%%%%%%%%%%%%%%%%%%%%%%%%%%%%%%%%%%%%%%%%%%%%%%%%%%
In this section we analyze the appearance of Berry's Quantum phase
in the phonon dynamics in the presence of a vortex in a superfluid. Now we consider in this geometry the appearance of Berry's quantum phase in phonons scattering by a vortex in a superfluid. The study of Berry's phase in the  superfluid context was investigated by several authors: Thouless, Ao and Niu\cite{prl:ao,prl:ao1} investigated the Berry phase in  vortex dynamics and pointed the connection  of the Magnus force with the Berry's phase; Schakel has investigated  Berry's quantum phase in the vortex dynamics in He-$3-A1$ superfluid\cite{epl:adr}. In the previous section we have demonstrated that phonons in a vortex background have a behavior similar to the photons in a background of the thick cosmic string. We use the approximation were $\frac{v_{s}^{2}}{c^{2}}<<1$.  Phonon propagation in the vortex background  is described by the Lorentzian equation
for the scalar field $g^{\mu\nu}\partial _{\mu}\partial_{\nu}\Psi(t,\rho,\phi,z)$. In the vortex metric this equation is given by
\begin{eqnarray}
  \label{b3}
  \{  \partial_{t}^{2} - \frac{1}{\rho}\partial_{\rho}(\rho\partial_{\rho})
 -\frac{1}{\rho^{2}}[(\beta \partial_{t}-\partial_{\phi})^{2}] +
  \partial_{z}^{2} \}\Psi(t,\rho,\phi,z)=0.
\end{eqnarray}
The background described by (\ref{spinning}) is time-independent and symmetric under
$z$ translations, therefore the solution of eq.(\ref{b3}) can be written as

\begin{equation}
  \Psi(t,\rho,\phi,z)=exp(-iEt)exp(ikz)\psi(\rho,\phi) \label{b4}
\end{equation}
where $E$ is the eigenvalue of energy and $k$ is the wave vector in the
z-direction. Using the Dirac phase factor method we can write
$\psi(\rho,\phi)$ as

\begin{eqnarray}
\psi(\rho,\phi)=exp\left(-i\int_{\phi_0}^{\phi}E\beta
d\phi\right)\psi_{0}(\rho,\phi) , \label{b5}
\end{eqnarray}
with $\psi_{0}(\rho,\phi)$ being a solution of the Klein-Gordon equation
in the ``space-time'' of a vortex in a superfluid.

Now we investigate Berry's phase for phonons in the vortex
background. For this case the geometric phase angle does depend on
the spectral label just as in the rotating cosmic string
case~\cite{mos}. Therefore, each different eigenmode, labeled by
$n$, acquires a different geometric phase, and as consequence the
appropriate treatment of this problem is obtained using the
non-Abelian generalization~\cite{mos} of  Berry's phase. In order
to compute this phase let us confine the quantum  system to a
perfectly reflecting  box such that the phonon  wave packet is
nonzero only in the interior of the box and is given by a
superposition of different phonon eigenfunctions. The vector that
localizes the box with respect to the vortex  is called  $\vec R$.
This vector is oriented from the origin of the coordinate system
(localized on the defect) to the center of the box. Call $R_{i}$
the components of $\vec R$, given by
$R_{i}=(R_{0},\phi_{0},z_{0})$ and such that $R_{0}>\beta$. 

In the absence of the vortex  the wave function
corresponding to the mode n is given by $\psi(\vec R - \vec x)$ where
$\vec x$ represents the coordinates of the phonons centered at $\vec R$.
When we consider the vortex , the wave function in the interior of the box is
obtained by use of the Dirac phase factor and is given by eq.(\ref{b5}). Let
the box be transported around a circuit $C$ threaded by the defect. Since the
space-time is axisymmetric we can transport the box along the Killing vector
field $R^{a}$.

Due to degeneracy of the energy eigenvalues, in order to compute
Berry's geometric phase, it is necessary to use the non-Abelian
version of the corresponding connection ~\cite{mos} given by

\begin{eqnarray}
 A_{n}^{IJ} = \langle\psi_{n}^{I}(\vec R - \vec x)|\nabla_{R}
\psi_{n}^{J}(\vec R - \vec x)\rangle,
\label{b5a}
\end{eqnarray}
where $I$ and $J$ stand for possible degeneracy labels.

The inner product in eq.(\ref{b5a})may be evaluated using by using the Dirac
phase factor

\begin{eqnarray}
&&\langle\psi^{I}(R_{i} - x_{i})|\nabla_{R}\psi_{n}^{J}(R_{i} - x_{i})\rangle=
\nonumber\\
&=&- i\oint_{\Sigma}dS \psi_{n}^{\ast I}(R_{i} - x_{i})
[- E_{n}\beta
\psi(R_{i} - x_{i}) \nonumber\\
&+& \nabla_{R}\psi_{n}^{J}(R_{i} - x_{i})].
\label{b5b}
\end{eqnarray}

The integrand  is calculated and the result is

\begin{eqnarray}
\langle\psi^{I}(R_{i} - x_{i})|\nabla_{R}\psi_{n}^{J}(R_{i} - x_{i})\rangle
=i E_{n}\beta \delta_{IJ} .
\label{b5c}
\end{eqnarray}

Berry's phase can be obtained from the expression(\ref{b5c}) and is given by

\begin{eqnarray}
\gamma(C)=  \frac{ E_{n}Nk}{c}  \label{b7},
\end{eqnarray}
where the labels $I,J$ and $\delta_{IJ}$ have been omitted. This
reproduces the results of Corrichi and Pierri~\cite{pie} and
Mostafazadeh\cite{mos} in the case of a spinning cosmic string.
The effect can be observed by an interference between the phonons
in the transported box and one in the box  that followed the
orbits of the Killing vector field. Note that the results
(\ref{b7}) are equal to the results of Stone~\cite{prb:stone} and
Volovik\cite{volo} for the analogue of the gravitational
Aharonov-Bohm effect for phonon scattering, in this way we
conclude that the gravitational Aharonov-Bohm in this system is a
particular case of Berry's quantum phase in this acoustical
analogue model.

%%%%%%%%%%%%%%%%%%%%%%%%%%%%%%%%%%%%%%%%%%%%%%%%%%%%%%%%%%%%%%%%%%%%%%%%%%%%%%%%%%%%%%%%%%%%%%%%%%%%%%%%%%%%%%%%%%%%%%%%%%%%%%%%%%%%%%%%%%%%%%%%%%%%%%%%%%%%%%%%%%%%%%%%%%%%%%%%%%%%%%%%
\section{Concluding Remarks}
%%%%%%%%%%%%%%%%%%%%%%%%%%%%%%%%%%%%%%%%%%%%%%%%%%%%%%%%%%%%%%%%%%%%%%%%%%%%%%%%%%%%%%%%%%%%%%%%%%%%%%%%%%%%%%%%%%%%%%%%%%%%%%%%%%%%%%%%%%%%%%%%%%%%%%%%%%%%%%%%%%%%%%%%%%%%%%%%%%%%%%%%%
In this work, we used an analogous model to describe a vortex in a
superfluid and have  used the holonomy matrix to show that this
system, i.e. phonons in the presence of a vortex,  presents the
Aharonov-Bohm effect similarly to particles in the presence of a
thick cosmic string. The Aharonov-Bohm phase has a contribution of both
translational and rotational holonomies. Far from the vortex only
translational holonomy contributes to the gravitational
Aharonov-Bohm effect, and the phonons in the presence of a vortex
line  behave like particles in the presence of  a spinning
massless cosmic string ~\cite{prd:bur}. We have studied  the
Berry's quantum phase for phonons in the presence superfluid
vortex in the point of view of the  Volovik  geometrical
description for superfluids, and the result obtained is similar to
the Aharonov-Bohm effect studied by Volovik and Stone. This fact
implies that the gravitational Aharonov-Bohm effect for phonons in
background of the superfluid vortex line is particular case of
Berry quantum phase.

\section{Acknowledgments}

One of us (GdA) would like to express his gratitude to
Professors Volovik and Letelier for helpful comments on the subject
of this paper. Financial supports from CNPq, CAPES(PROCAD) and UERJ is
gratefully acknowledged.


\begin{thebibliography}{7}

\bibitem{bjp:moraes}
Fernando Moraes,
 Brazilian Journal of Physics, {\bf 30} 2(2000).
\bibitem{prl:gar} L. J. Garay, J. R. Anglin, J. I. Cirac, and P. Zoller Phys. Rev. lett. {\bf 85}, 4643 (2000).
\bibitem{pra:gar}L. J. Garay, J. R. Anglin, J. I. Cirac and P. Zoller Phys. Rev. {\bf 63} 023611 (2001).
\bibitem{prl:unruh}W. G. Unruh, Phys. Rev. {\bf D 51}, 2827 (1995).
\bibitem{prd:unruh}
W. G. Unruh,
Phys. Rev. D, {\bf 51} 6(1995).
\bibitem{prd:jac}T. Jacobson, Phys. Rev. {\bf D 44}, 1731 (1991).
\bibitem{prl:visser}M. Visser, Phys. Rev. Lett. {\bf 80}, 3436 (1998).
\bibitem{cqg:visser}
Matt Visser
Class. Quant. Grav. {\bf 15} 1767 (1998).
\bibitem{volo}
G. E. Volovik,
JETP LETT, {\bf 67} (11) (1998).
\bibitem{boo} G. E. Volovik {\it The Universe in a Helium Droplet}( Oxford University Press)2003.
\bibitem{leon}Ulf Leonhardt, Phys. Rev. {\bf A 65}, 043818 (2002)
\bibitem{brev} I. Brevik et al., Phys. Rev. {\bf D 65}, 024005 (2002).
\bibitem{prs:hall}H. E. Hall and Vinen, Proc. Roy. Soc. {\bf A238}, 204,(1956).
\bibitem{jetp:ior} S. V. Iordanskii, Zh. \'Eksp. Teor. Fiz. {\bf 49}, 225 (1965)[Sov. Phys. JETP {\bf 22}, 160 (1966)].
\bibitem{jetp:sonin} E. B. Sonin,  Zh. \'Eksp. Teor. Fiz. {\bf 69},921 (1975)[Sov. Phys. JETP {\bf 42}, 469 (1976)].
\bibitem{prb:sonin} E. B. Sonin, Phys Rev. {\bf B 55},485 (1997).
\bibitem{prb:stone}Michael Stone, Phys Rev , {\bf B 61} 11780  (2000).
\bibitem{ejp:berry} M. V. Berry, R. G. Chambers, M. D. Large, C. Upstill, and J. C. Walmsley, Eur. J.  Phys, {\bf 1},154 (1980).
\bibitem{prl:ao}P. Ao, D. J. Thouless, Phys. Rev. Lett. {\bf 70} 2158 (1993).
\bibitem{prl:ao1} D. J. Thouless, P. Ao, and Q. Niu, Phys. Rev. Lett. {\bf 76} 3758 (1996).
\bibitem{prl:wex} C. Wexler, Phys. Rev. Lett. {\bf 79} 1321 (1997).
\bibitem{prb:wex}C. Wexler and D. J. Thouless, Phys. Rev. {\bf B 58},8897 (1998).
\bibitem{prb:rhe} D. J. Thouless, M. R. Geller,W. F. Vinen, J. Y. Fortin and S. W. Rhee, Phy Rev {\bf B  63} 224504 (2001) .
\bibitem{condm:sonin} E. Sonin Cond-mat 0104221.
\bibitem{pr:aha}  Y. Aharonov and D. Bohm, Phys. Rev. {\bf 115, }485 (1959).
\bibitem{jphys:vile}  L.H. Ford and A. Vilenkin, J. Phys. {\bf A14}, 2353 (1981).
\bibitem{prd:val} V. B. Bezerra, Phys. Rev. {\bf D35}, 2031 (1987).
\bibitem{an:val}V. B. Bezerra, Ann. Phys. (NY) {\bf 203} 392 (1990)
\bibitem{nuo:law}  J. K. Lawrence, D. Leiter and G. Szamosi, Nuovo Cimento {\bf 17B%
}, 113 (1973).
\bibitem{nuo:fro}  V. P. Frolov and V. D. Skarzhinsky, Nuovo Cimento  {\bf 99B},
67 (1987).
\bibitem{prl:visfis} M. Visser and  U. R. Fischer, Phys. Rev. Lett. {\bf 88}, 110201 (2002).
\bibitem{an:visfis}M. Visser and  U. R. Fischer, Ann. Phys. (NY) {\bf 304} 22 (2003)
\bibitem{pla:anandan} J. Anandan, Phys. Lett. A {\bf 195} 284 (1994).
\bibitem{cqg:solen} R. A. Puntigam and  H.Soleng, Classic Quant. Grav {\bf 14},1129 (1997).
\bibitem{cqg:garcia} L. C. Garcia de Andrade, Class. Quant. Grav. {\bf 16} 1583 (1999).
\bibitem{cqg:tod}K. P. Tod, Class. Quant. Grav. {\bf 11} 1331 (1994).

\bibitem{pla:kohler} Christopher Kohler, Phys. Letts A  {\bf 237} (4-5): 195 (1998).

\bibitem{petit}R. J. Petit Genn. Rel. and Grav. {\bf 18} 5 (1986).
\bibitem{prd:bur}C.J. C. Burgers Phys Rev. {\bf D 32} 504 (1985).
\bibitem{cqg:vic} J. A. G. Vickers, Class.Quantum Grav. {\bf 4} 1 (1987).
\bibitem{epl:adr} A. M. J. Schakel, Europhys. Lett {\bf 10} 159 (1989).
\bibitem{mos} A. Mostafazadeh, J. Phys. {\bf A31}, 7829 (1998).
\bibitem{pie} A. Corichi and M. Pierri, Phys. Rev. {\bf D51}, 5870 (1995).
\end{thebibliography}
\end{document}